\documentclass[a4paper,onecolumn,11pt,unpublished]{quantumarticle}

\pdfoutput=1

\usepackage[numbers,sort&compress]{natbib}
\usepackage[utf8]{inputenc}
\usepackage[english]{babel}
\usepackage[T1]{fontenc}
\usepackage{amsmath,amssymb,bm}
\usepackage{graphics}
\usepackage{dsfont}

\newcommand{\ket}[1]{\left|{#1}\right\rangle}
\newcommand{\bra}[1]{\left\langle{#1}\right|}
\newcommand{\braket}[2]{\left\langle{#1}|{#2}\right\rangle}
\newcommand{\braketa}[2]{\left\langle \left.{#1}\right|{#2}\right\rangle}
\newcommand{\braketb}[2]{\left\langle {#1}\left|{#2}\right.\right\rangle}

\newcommand{\dbar}{d\hspace*{-0.08em}\bar{}\hspace*{0.1em}}
\newcommand{\Dcirc}{\mathcal{D}^{\circ}}

\begin{document}

\title{Wigner functional theory for quantum optics}

\author{Filippus S. Roux}
\email{froux@nmisa.org}
\affiliation{National Metrology Institute of South Africa, Meiring Naud{\'e} Road, Brummeria 0040, Pretoria, South Africa}
\orcid{0000-0001-9624-4189}

\author{Nicolas Fabre}
\email{nicolas.fabre@univ-paris-diderot.fr}
\affiliation{Laboratoire Mat\'eriaux et Ph\'enom\`enes Quantiques, Sorbonne Paris Cit\'e, Universit\'e de Paris, CNRS UMR 7162, 75013 Paris, France}

\begin{abstract}
Using the quadrature bases that incorporate the spatiotemporal degrees of freedom, we develop a Wigner functional theory for quantum optics, as an extension of the Moyal formalism. Since the spatiotemporal quadrature bases span the complete Hilbert space of all quantum optical states, it does not require factorization as a tensor product of discrete Hilbert spaces. The Wigner functions associated with such a space become functionals and operations are expressed by functional integrals --- the functional version of the star product. The resulting formalism enables tractable calculations for scenarios where both spatiotemporal degrees of freedom and particle-number degrees of freedom are relevant. To demonstrate the approach, we compute examples of Wigner functionals for a few well-known states and operators.
\end{abstract}

\maketitle


\section{\label{intro}Introduction}

Quantum information technology promises to provide secure communication \cite{gisin}, more accurate measurements \cite{giovannetti} and more efficient computations \cite{steane}, among other benefits. However, quantum states are often fragile. The purity and coherence of such states, for instance, are easily lost when such states interact with the environment \cite{schlosshauer}.

To increase the information capacity of quantum systems \cite{hausladen,lloyd,holevo} and to improve the security in quantum cryptography \cite{bp2000,cerf,scarani}, the states are often prepared in higher dimensional Hilbert spaces. An example is the spatial modes of photons, such as orbital angular momentum (OAM) modes \cite{allen,cryptotwist}. They represent an infinite dimensional Hilbert space. Applications that use such higher dimensional Hilbert spaces are usually implemented in terms of individual photons encoded in terms of their spatial degrees of freedom. Losses and stray photons tend to reduce the fidelity, purity or signal-to-noise ratio in quantum states, slowing down the rate at which such systems can operate \cite{waks}.

A way to overcome the losses and noise issues is to prepare multiphoton states that also incorporate different spatial modes. Such quantum systems are defined in terms of both their spatiotemporal degrees of freedom and particle-number degrees of freedom \cite{sanders}. They are often rather complex and difficult to analyze. One approach is to duplicate the operator formalism for a single-mode multi-particle system several times to handle several discrete modes \cite{contvar1,contvar2}. The result is best applied in cases of Gaussian states that can be represented in terms of a few discrete spatial modes \cite{weedbrook}.

In a different development, started during the Second World War, it was independently shown by Groenewold \cite{groenewold} and Moyal \cite{moyal} that quantum mechanics can be successfully formulated without operators. This formulation of quantum mechanics in phase space \cite{psqm} represents the states and operators by functions of phase space variables (analogues to position and momentum for the harmonic oscillator). Examples of such functions are the quasi-probability distributions that include the Glauber-Sudarshan $P$-distribution \cite{glauber,sudarshan}, the Husimi $Q$-distribution \cite{husimi} and the Wigner distribution \cite{wigner}. Products of operators are represented by so-called {\em star products} of the phase space functions. The Moyal formulation was shown to reproduce all the uncertainty relations associated with quantum mechanics.

One of the challenges initially encountered with the Moyal formulation was how to incorporate other degrees of freedom (apart from the particle-number degrees of freedom) into the formulation. In the case of spin (and other internal symmetries), the problem was overcome with the aid of the Stratonovich-Weyl correspondence \cite{stratonovich,brif,tilma}. For the spatial degrees of freedom, one can use a similar approach \cite{rundle} or other approaches (see for example \cite{krumm}), but these again lead to a finite set of discrete spatial modes (and often tend to return to an operator-based approach).

Recently, the spatiotemporal degrees of freedom and the particle-number degrees of freedom were combined into one comprehensive Hilbert space for all quantum optical states \cite{stquad,stquaderr}, by introducing {\em spatiotemporal quadrature bases} that are generalizations of the quadrature bases associated with only the particle-number degrees of freedom. This approach was subsequently used to investigate the evolution of arbitrary multiphoton states propagating through turbulence \cite{ifpe} and to investigate the amount of entanglement in a parametric down-converted state when all the degrees of freedom are included \cite{entpdc}. However, a full description of the approach has not been published yet.

The purpose of this article is to provide such a description. Guided by the Moyal formalism, we use the spatiotemporal quadrature bases to develop a formalism that incorporates both the spatiotemporal degrees of freedom and the particle-number degrees of freedom. For this purpose, we choose the Wigner distribution, since it is naturally related to the quadrature bases. (However, the resulting formalism can be used with any of the other quasi-probability distributions.) In this approach, these quadrature bases are used to generalize the standard Wigner distributions to become {\em Wigner functionals}. The development parallels the normal theory of Wigner functions (and of the Moyal formalism), showing that most properties can be carried over to the functional formalism, except that analyses now tend to involve functional integrals. Though the expressions may appear familiar, the resemblance is deceptive in that it now incorporates all the spatiotemporal degrees of freedom. While the current development is done in the context of quantum optics, similar developments have been done in the context of quantum field theory \cite{Mrowczynski_Wigner_1994,Mrowczynski_Wigner_2013,recent}.

The involvement of functional integrals in the new formalism may create the impression that any analysis that is done with this formalism would be severely complex and often intractable. It is true that, apart from some special cases, one can evaluate such functional integrals only when the integrand is in the form of a Gaussian functional. However, with the aid of auxiliary variables, source terms and generating functionals, it is often possible to represent the quantum states and operations in terms of Gaussian functionals, even if the original expressions are not of that form.

To demonstrate its usefulness, we use the formalism to compute the Wigner functionals for a few well-known states and operators. For example, we compute a generating functional for the Wigner functionals of fixed-spectrum Fock states. The term {\em fixed-spectrum} indicates that the spatiotemporal degrees of freedom of all the photons in the state are represented by the same spectrum of plane waves. Although the Wigner functionals of Fock states are not in Gaussian form, their generating functional is in Gaussian form and can therefore be used in calculations involving functional integrals.

The functional phase space introduced here can be considered as the ``mother'' of all phase spaces for quantum optics in that it allows one to recover all those that are studied in quantum optics. For example, by restricting the spatiotemporal degrees of freedom to a single spectral function, the functional phase space reduces to the phase space that only represents the particle-number degrees of freedom.

The paper is organized as follows. In Sec.~\ref{quadb}, we review the spatiotemporal quadrature bases, together with some background on other aspects that we need in the rest of the paper. The definition of the Wigner functionals and related quantities are discussed in Sec.~\ref{wfth}. Some examples of Wigner functionals are computed in Sec.~\ref{voorb}. We provide a discussion in Sec.~\ref{disc} and end with conclusions in Sec.~\ref{concl}.


\section{\label{quadb}Background}

\subsection{\label{fmqb}Eigenstates of quadrature operators}

The quadrature bases in terms of which the Wigner functionals for quantum optics is formulated, are obtained as eigenstates of the wave vector dependent quadrature operators:
\begin{align}
\begin{split}
\hat{q}_s(\mathbf{k})\ket{q} & = \ket{q} q_s(\mathbf{k}) , \\
\hat{p}_s(\mathbf{k})\ket{p} & = \ket{p} p_s(\mathbf{k}) .
\end{split}
\label{eieqpfm}
\end{align}
Here, $\mathbf{k}$ represents the three-dimensional wave vector and the subscript $s$ is the spin index. These quadrature operators are directly defined in terms of the standard ladder operators $\hat{a}^{\dag}_s(\mathbf{k})$ and $\hat{a}_s(\mathbf{k})$ used for the quantization of the electromagnetic field:
\begin{align}
\begin{split}
\hat{q}_s(\mathbf{k}) & = \frac{1}{\sqrt{2}} \left[\hat{a}_s(\mathbf{k})+\hat{a}^{\dag}_s(\mathbf{k})\right] , \\
\hat{p}_s(\mathbf{k}) & = \frac{-i}{\sqrt{2}} \left[\hat{a}_s(\mathbf{k})-\hat{a}^{\dag}_s(\mathbf{k})\right] .
\end{split}
\label{quadopdef}
\end{align}
The ladder operators obey a Lorentz covariant commutation relation, given by
\begin{equation}
\left[\hat{a}_s(\mathbf{k}_1), \hat{a}_r^{\dag}(\mathbf{k}_2)\right] = (2\pi)^3 \omega_1\delta_{s,r}\delta(\mathbf{k}_1-\mathbf{k}_2) ,
\label{comaafm}
\end{equation}
where $\omega_1=c|\mathbf{k}_1|$ is the angular frequency, given in terms of the vacuum dispersion relation, $\delta_{s,r}$ is the Kronecker delta for the spin indices and $\delta(\mathbf{k}_1-\mathbf{k}_2)$ is a three-dimensional Dirac $\delta$ function for the wave vectors. The equivalent Lorentz covariant commutation relation for the quadrature operators reads
\begin{equation}
\left[ \hat{q}_s(\mathbf{k}_1), \hat{p}_r(\mathbf{k}_2) \right] = i (2\pi)^3 \omega_1 \delta_{s,r} \delta(\mathbf{k}_1-\mathbf{k}_2) .
\label{compqfm}
\end{equation}
Although the two quadrature operators $\hat{q}_s(\mathbf{k})$ and $\hat{p}_s(\mathbf{k})$ in \eqref{eieqpfm} are unique operator-valued functions of the wave vector, the real-valued eigenvalue functions $q_s(\mathbf{k})$ and $p_s(\mathbf{k})$ are not unique --- there are an infinite number of them. Indeed, $q_s,p_s \in L^2\{\mathbb{R}^3\}$. Moreover, for each eigenvalue function, $q_s(\mathbf{k})$ or $p_s(\mathbf{k})$, there is a unique eigenstate, $\ket{q}$ or $\ket{p}$, which is associated with the function as a whole and not with particular function values of the eigenvalue function. For that reason, an eigenstate does not explicitly depend on the value of the wave vector.

To simplify notation, we shall neglect the spin degrees of freedom and not display the spin indices in the remainder of this paper. It is nevertheless straight-forward to reintroduce them if necessary.

The eigenstates in \eqref{eieqpfm} can be created from the vacuum
\begin{align}
\begin{split}
\ket{q} & = \hat{a}_q^{\dag} \ket{\text{vac}} , \\
\ket{p} & = \hat{a}_p^{\dag} \ket{\text{vac}} ,
\end{split}
\end{align}
with operators given by
\begin{align}
\begin{split}
\hat{a}_q^{\dag} & = \pi^{-\Omega/4} \exp\left(-\tfrac{1}{2}\|q\|^2+\hat{a}_{Q}^{\dag}-\hat{a}_{R}^{\dag}\right) , \\
\hat{a}_p^{\dag} & = 2^{\Omega/2}\pi^{\Omega/4} \exp\left(-\tfrac{1}{2}\|p\|^2+i\hat{a}_{P}^{\dag}+\hat{a}_{R}^{\dag}\right) ,
\end{split}
\label{defqpskep}
\end{align}
where
\begin{align}
\begin{split}
\hat{a}_{Q}^{\dag} & = \sqrt{2} \int \hat{a}^{\dag}(\mathbf{k}) q(\mathbf{k})\ \dbar k , \\
\hat{a}_{P}^{\dag} & = \sqrt{2} \int \hat{a}^{\dag}(\mathbf{k}) p(\mathbf{k})\ \dbar k , \\
\hat{a}_{R}^{\dag} & = \frac{1}{2}\int \hat{a}^{\dag}(\mathbf{k}) \hat{a}^{\dag}(\mathbf{k})\ \dbar k ,
\end{split}
\label{defqra}
\end{align}
and,
\begin{equation}
\|f\|^2 \equiv \int \left|f(\mathbf{k})\right|^2\ \dbar k ,
\label{normfunk}
\end{equation}
for $f:\mathbb{R}^3 \to \mathbb{C}$. The quantity $\Omega$ in \eqref{defqpskep} is defined as
\begin{equation}
\Omega \equiv \int \delta(0)\ \text{d}^3 k ,
\label{defomega}
\end{equation}
and (within the current context) it represents the cardinality of a countable infinite set $\Omega=\aleph_0$. The integration measures in \eqref{defqpskep}--\eqref{normfunk} and below are given in terms of a simplified notation:
\begin{equation}
\dbar k \equiv \frac{\text{d}^3 k}{(2\pi)^3 \omega} .
\end{equation}
Note that all the wave vector dependencies are integrated out in \eqref{defqpskep}, so that the elements of the quadrature bases do not explicitly depend on the wave vector.

\subsection{\label{orto}Orthogonality and completeness}

The quadrature bases obey orthogonality conditions \cite{stquad,stquaderr}, expressed in terms of Dirac $\delta$ functionals
\begin{align}
\begin{split}
\braket{q}{q'} & = \delta[q-q'] , \\
\braket{p}{p'} & = (2\pi)^{\Omega} \delta[p-p'] .
\end{split}
\label{ortoqp}
\end{align}
The square brackets indicate that the quantity is a functional (a function of functions), where $q$ and $p$ represent functions. It depends on the entire functions and not on particular function values of those functions. For that reason, the quantity does not explicitly depend on $\mathbf{k}$ and we do not show the arguments of the functions inside the square brackets.

The completeness conditions for the spatiotemporal quadrature bases \cite{stquad,stquaderr} are represented as functional integrals
\begin{align}
\begin{split}
\int \ket{q} \bra{q}\ \mathcal{D}[q] & = \mathds{1} , \\
\int \ket{p} \bra{p}\ \Dcirc [p] & = \mathds{1} ,
\end{split}
\label{volqp}
\end{align}
where $\mathds{1}$ is the identity operator for the entire Hilbert space of all quantum optical states. The functional measures in \eqref{volqp} run over all finite-energy real-valued functions. The measure for the integral over $p$ incorporates a normalization constant:
\begin{equation}
\Dcirc [p] \equiv \mathcal{D}\left[\frac{p}{2\pi}\right] = \frac{1}{(2\pi)^{\Omega}}\ \mathcal{D}[p] .
\end{equation}

\subsection{\label{afbind}Notation for contraction}

Since these functional integrals can, apart from some special cases, only be evaluated when their integrands are in Gaussian form, one can simplify the notation. The Gaussian form implies an exponential function with an argument consisting of integrals over some degrees of freedom, typically the three-dimensional wave vectors. The integrands of these integrals are products of functions of the wave vectors. There may be multiple distinct wave vectors that are being integrated. Usually, a given wave vector would appear exactly twice as arguments of functions in each term, thus connecting a pair of functions in the term. We denote such a connection by a binary operator $\diamond$ and we call it a $\diamond$-{\em contraction}.

As an example, we introduce the following notation for the inner-product between two functions
\begin{equation}
\langle f,g\rangle \equiv \int f^*(\mathbf{k}) g(\mathbf{k})\ \dbar k \equiv f^*\diamond g .
\label{binnespek}
\end{equation}
If there is a kernel function involved, we have
\begin{equation}
f^*\diamond B \diamond g \equiv \int f^*(\mathbf{k}) B(\mathbf{k},\mathbf{k}') g(\mathbf{k}')\ \dbar k\ \dbar k' .
\end{equation}
Note that it is not equivalent to
\begin{equation}
f^*\diamond T \diamond g \neq \int f^*(\mathbf{k}) T(\mathbf{k}) g(\mathbf{k})\ \dbar k .
\end{equation}

\subsection{\label{fopi}Fourier integrals}

An important quantity is the overlap $\braket{q}{p}$, which reads
\begin{equation}
\braket{q}{p} = \exp\left[ i \int q(\mathbf{k}) p(\mathbf{k}) \dbar k \right] \equiv \exp(i q\diamond p) .
\label{oorvqp}
\end{equation}
It appears when expressions are converted from one quadrature basis into another mutually unbiased quadrature basis and thus represents the kernel of a functional Fourier transform. These Fourier transforms indicate that the quadrature bases are related by functional Fourier transforms. As a result, one can express one in terms of the other as
\begin{align}
\begin{split}
\ket{p} & = \int \ket{q} \exp(i q\diamond p)\ \mathcal{D}[q] , \\
\ket{q} & = \int \ket{p} \exp(-i q\diamond p)\ \Dcirc [p] .
\end{split}
\label{pqftqp}
\end{align}
Using the expressions of the eigenvalue equations in \eqref{eieqpfm}, the Fourier relationships allow the quadrature operators to be represented in their dual bases by {\em functional derivatives}
\begin{align}
\begin{split}
\hat{p}(\mathbf{k}) & = \int \ket{q} \left[-i \frac{\delta}{\delta q(\mathbf{k})}\right] \bra{q}\ \mathcal{D}[q] , \\
\hat{q}(\mathbf{k}) & = \int \ket{p} \left[i \frac{\delta}{\delta p(\mathbf{k})}\right] \bra{p}\ \Dcirc [p] .
\end{split}
\label{qpoper}
\end{align}
The operation of a functional derivative is defined by
\begin{equation}
\frac{\delta}{\delta f(\mathbf{k})} f(\mathbf{k}') = (2\pi)^2\omega\delta(\mathbf{k}-\mathbf{k}') \equiv \mathbf{1} .
\label{funcdif}
\end{equation}

One can also use the definitions of the quadrature bases in terms of their dual bases, given in \eqref{pqftqp}, to define shift operators
\begin{align}
\begin{split}
\ket{q-q_0} & = \exp(i q_0\diamond\hat{p})\ \ket{q} , \\
\ket{p-p_0} & = \exp(-i \hat{q}\diamond p_0)\ \ket{p} .
\end{split}
\label{skuifqp}
\end{align}
These shift operators will come in handy later.

\subsection{\label{dirac}Integrals for Dirac $\delta$ functionals}

Using \eqref{ortoqp}, \eqref{volqp}, and \eqref{oorvqp}, one can show that
\begin{align}
\begin{split}
\int \exp(i q_1\diamond p-i q_2\diamond p)\ \Dcirc [p] & = \delta[q_1-q_2] , \\
\int \exp(-i q\diamond p_1+i q\diamond p_2)\ \mathcal{D}[q] & = (2\pi)^{\Omega} \delta[p_1-p_2] .
\end{split}
\label{diracft}
\end{align}
Combining these integrals and converting the real-valued {\em field variables} into complex-valued field variables, given by
\begin{equation}
\alpha(\mathbf{k}) = \frac{1}{\sqrt{2}}\left[q(\mathbf{k})+i p(\mathbf{k})\right] ,
\label{kohreim}
\end{equation}
one obtains
\begin{equation}
\int \exp\left(\alpha_0^*\diamond\alpha-\alpha^*\diamond\alpha_0\right)\ \Dcirc[\alpha] = (2\pi)^{\Omega} \delta[\alpha_0] ,
\label{adirac}
\end{equation}
where
\begin{equation}
\Dcirc [\alpha] \equiv \mathcal{D}[q]\ \Dcirc [p] ,
\label{alfamaat}
\end{equation}
and
\begin{equation}
\delta[\alpha_0] \equiv \delta[q_0]\ \delta[p_0] .
\end{equation}

The complex-valued function $\alpha(\mathbf{k})$, defined in \eqref{kohreim}, can serve different purposes. It can be regarded as an independent field variable in the context of functional expressions and thus can become the integration variable in functional integrals. It can also serve as a {\em parameter function}, representing for instance the spectral function in fixed-spectrum coherent states, considered below. Such parameter functions can also be turned into integration (fields) variables for functional integrals.

\subsection{\label{gaussint}Gaussian integrals}

The generic functional integral with an integrand in isotropic Gaussian form can be evaluated to give
\begin{equation}
\int \exp \left(-\alpha^*\diamond K\diamond\alpha-\alpha^*\diamond\xi-\zeta^*\diamond\alpha\right)\ \Dcirc [\alpha]
 = \frac{\exp\left( \zeta^*\diamond K^{-1}\diamond\xi \right)}{\det\{K\}} ,
\label{genfint}
\end{equation}
where $K$ is an invertible kernel, and $\xi$ and $\zeta$ are arbitrary complex functions. Invertibility implies that the kernel must have an inverse $K^{-1}$, such that
\begin{equation}
K\diamond K^{-1} = \int K(\mathbf{k}_1,\mathbf{k}) K^{-1}(\mathbf{k},\mathbf{k}_2)\ \dbar k = \delta(\mathbf{k}_1-\mathbf{k}_2) .
\label{invkern}
\end{equation}
The functional determinant $\det\{K\}$ can be expressed as
\begin{equation}
\det\{K\} = \exp[\text{tr}\{\ln_{\diamond}(K)\}] ,
\label{detdef}
\end{equation}
where $\ln_{\diamond}(\cdot)$ is defined as the inverse of
\begin{equation}
\exp_{\diamond}(H) \equiv \mathbf{1} + \sum_{n=1}^{\infty} \frac{1}{n!} (H)^{\diamond n} ,
\label{expdef}
\end{equation}
for an arbitrary kernel function $H(\mathbf{k}_1,\mathbf{k}_2)$, and the trace of such a kernel is given by
\begin{equation}
\text{tr}\{H\} \equiv \int H(\mathbf{k},\mathbf{k})\ \dbar k .
\label{trdef}
\end{equation}

\subsection{\label{fsfock}Fixed-spectrum Fock states}

The {\em fixed-spectrum} Fock states are defined as
\begin{equation}
\ket{n_F} = \frac{1}{\sqrt{n!}}\left({\hat{a}^{\dag}}_F\right)^n \ket{\text{vac}} ,
\label{fsfsdef}
\end{equation}
in terms of fixed-spectrum creation operators, defined by
\begin{equation}
{\hat{a}^{\dag}}_F \equiv \int \hat{a}^{\dag}(\mathbf{k}) F(\mathbf{k})\ \dbar k .
\label{fsskepabs}
\end{equation}
The angular spectrum (or Fourier domain wave function) $F(\mathbf{k})$ that parameterizes a fixed-spectrum Fock state is normalized:
\begin{equation}
\|F\|^2=\int \left| F(\mathbf{k}) \right|^2\ \dbar k = 1 .
\label{norm1}
\end{equation}
It ensures that the fixed-spectrum ladder operators obey a simple commutation relation $[\hat{a}_F,{\hat{a}_F}^{\dag}]=1$ and that each fixed-spectrum Fock state is normalized $\braket{n_F}{n_F}=1$. The inner-product between Fock states with different spectra reads
\begin{equation}
\braket{m_F}{n_G} = \delta_{mn} (\langle F,G \rangle)^n ,
\label{infockf}
\end{equation}
where $\langle F,G \rangle$ is defined in \eqref{binnespek}. When the annihilation operator in the momentum basis is applied to the fixed-spectrum Fock states, we obtain
\begin{equation}
\hat{a}(\mathbf{k}) \ket{n_F} = \ket{(n-1)_F} F(\mathbf{k}) \sqrt{n} .
\label{akopnfs}
\end{equation}
The fixed-spectrum Fock states are eigenstates of the number operator
\begin{equation}
\hat{n} \equiv \int \hat{a}^{\dag}(\mathbf{k}) \hat{a}(\mathbf{k})\ \dbar k .
\label{numopk}
\end{equation}
Using \eqref{akopnfs}, one can show that
\begin{equation}
\hat{n} \ket{n_F} = \ket{n_F} n .
\end{equation}

\subsection{\label{fskoh}Fixed-spectrum coherent states}

The fixed-spectrum coherent states are defined as eigenstates of the standard (wave vector dependent) annihilation operator
\begin{equation}
\hat{a}(\mathbf{k}) \ket{\alpha_F} = \ket{\alpha_F} \alpha(\mathbf{k}) ,
\label{akopkohm}
\end{equation}
where the eigenvalue function $\alpha(\mathbf{k})$ is an arbitrary finite-energy complex-valued spectral function $\alpha:\mathbb{R}^3 \to \mathbb{C}$. The fixed-spectrum coherent states $\ket{\alpha_F}$ do not explicitly depend on $\mathbf{k}$. There exists a unique fixed-spectrum coherent state for every spectral function $\alpha(\mathbf{k})$. The concept of a fixed-spectrum coherent states is not new \cite{combescure_coherent_2012}. They have been used in various contexts, for instance in condensed matter theory to calculate Green functions \cite{sachdev_quantum_2001}.

The subscript $F$ in $\ket{n_F}$, $\ket{\alpha_F}$ and ${\hat{a}^{\dag}}_F$ is a reminder that the state or operator is parameterized in terms of a fixed spectrum and should not necessarily be seen as a label for the associated complex-valued function. The latter is thus represented as $\alpha(\mathbf{k})$ and not $\alpha_F(\mathbf{k})$. Later, where we use different fixed-spectrum coherent states in the same expression, we will use the parameter functions to label the coherent states, instead of $\alpha_F$.

The fixed-spectrum coherent states can be expressed in terms of functional displacement operators given by
\begin{equation}
\hat{D}[\alpha_F] \equiv \exp\left(\alpha\diamond\hat{a}^{\dag}-\alpha^*\diamond\hat{a}\right) ,
\label{verplfs}
\end{equation}
which are generalizations of the multimode displacement operator \cite{barnett_methods_2002}. The inner-product between different fixed-spectrum coherent states can be derived from their displacement operators and reads
\begin{equation}
\braket{\alpha_F}{\beta_G} = \exp \left(-\tfrac{1}{2}\|\alpha\|^2-\tfrac{1}{2}\|\beta\|^2+\langle\alpha,\beta\rangle \right) .
\label{fsinprodf}
\end{equation}
As a consequence, it follows that the inner-product between a fixed-spectrum coherent state and the vacuum state is
\begin{equation}
\braket{\alpha_F}{\text{vac}} = \exp \left(-\tfrac{1}{2}\|\alpha\|^2\right) .
\label{fsinvac}
\end{equation}

Although not orthogonal, the fixed-spectrum coherent states resolve the identity operator. The completeness condition for the fixed-spectrum coherent states reads \cite{stquaderr}
\begin{equation}
\mathds{1} = \int \ket{\alpha} \bra{\alpha}\ \Dcirc[\alpha] .
\label{volkoh}
\end{equation}

\subsection{\label{qfskoh}Quadrature representation of coherent states}

To expand fixed-spectrum coherent states in terms of the spatiotemporal quadrature bases, we use the operators defined in \eqref{defqra} and employ the eigenstate property of the coherent states in \eqref{akopkohm}. As a result, we obtain
\begin{align}
\begin{split}
\hat{a}_{Q}\ket{\alpha_F} & = \ket{\alpha_F} \sqrt{2} q\diamond\alpha_0 , \\
\hat{a}_{R}\ket{\alpha_F} & = \ket{\alpha_F} \tfrac{1}{2} \alpha_0\diamond\alpha_0 ,
\end{split}
\end{align}
where $\alpha_0$ represents the complex-valued parameter function of the fixed-spectrum coherent state. Therefore,
\begin{align}
\braket{q}{\alpha_F} = & \pi^{-\Omega/4} \bra{\text{vac}} \exp\left(-\tfrac{1}{2} \|q\|^2+\hat{a}_{Q}-\hat{a}_{R}\right) \ket{\alpha_F} \nonumber \\
 = & \pi^{-\Omega/4}\exp\left(-\tfrac{1}{2} \|q\|^2 -\tfrac{1}{2} \|\alpha_0\|^2 + \sqrt{2} q\diamond\alpha_0-\tfrac{1}{2} \alpha_0\diamond\alpha_0\right) .
\label{oorvqkoh}
\end{align}
If we express $\alpha_0(\mathbf{k})$ in terms of its real and imaginary parts, as in \eqref{kohreim}, we obtain
\begin{equation}
\braket{q}{\alpha_F} = \pi^{-\Omega/4} \exp\left[i p_0\diamond \left(q-\tfrac{1}{2}q_0\right) -\tfrac{1}{2} \|q-q_0\|^2\right] .
\label{oorvqkoh0}
\end{equation}

\section{\label{wfth}Wigner functional theory}

Here, we develop the formalism for Wigner functionals in quantum optics. To avoid cluttering the notations, we proceed to neglect the spin indices. However, we emphasize that one can incorporate spin into the formalism when necessary. Therefore, the resulting formalism represents all the degrees of freedom of quantum optics.

The expressions that are obtained resemble those in the standard formalism that only involves the particle-number degrees of freedom. The expressions we obtain here often look the same. However, we emphasize that due to the functional nature of these expressions, they also incorporate all the spatiotemporal degrees of freedom.

\subsection{\label{wigfnaldef}Definition of the Wigner functional}

The generic definition of a {\em Wigner functional} is
\begin{equation}
W[q,p] \equiv \int \bra{q+\tfrac{1}{2}x}\hat{A}\ket{q-\tfrac{1}{2}x} \exp(-i p\diamond x)\ \mathcal{D}[x] ,
\label{wigdef}
\end{equation} 
where $\hat{A}$ is an operator on the Hilbert space of all quantum optical states, incorporating both particle-number degrees of freedom and spatiotemporal degrees of freedom. The square brackets in $W[q,p]$ indicate that the quantity is a functional. While $q$ and $p$ represent two real-valued functions, we'll eventually combine them into a single complex-valued function $\alpha$ and represent the Wigner functional as $W[\alpha]$. As in the case of Wigner functions, which only represent the particle-number degrees of freedom, Wigner functionals, which represent both the particle-number degrees of freedom and the spatiotemporal degrees of freedom, can be negative and thus need to be interpreted as quasi-probability distributions.

If the operator is a density operator $\hat{\rho}$, the resulting Wigner functional represents a quantum state. The density operator can also be represented as a density `matrix', which we refer to as a {\em density functional}
\begin{equation}
\rho\left[q+\tfrac{1}{2}x,q-\tfrac{1}{2}x\right] \equiv \bra{q+\tfrac{1}{2}x}\hat{\rho}\ket{q-\tfrac{1}{2}x} .
\end{equation}
In the case of a pure state, the density functional becomes a product of a {\em wave functional} $\psi[q]=\braket{q}{\psi}$ and its complex conjugate
\begin{equation}
\rho\left[q+\tfrac{1}{2}x,q-\tfrac{1}{2}x\right] = \psi\left[q+\tfrac{1}{2}x\right]\psi^*\left[q-\tfrac{1}{2}x\right] .
\end{equation}

One can convert the calculation of the Wigner functional into a purely operator-based calculation. For this purpose, we use the shift operators defined in \eqref{skuifqp} and pull the Fourier kernel through the quadrature basis elements to convert it into operators. When these operators on either side of the density operator are combined with the aid of the Baker-Campbell-Hausdorff formula
the result reads
\begin{equation}
W[q,p] = 2^{\Omega} \text{tr} \left\{ \hat{\rho} \hat{D}[q,p] \hat{\Pi} \hat{D}^{\dag}[q,p] \right\} ,
\end{equation}
where $\hat{D}[q,p]$ is the displacement operator, defined in \eqref{verplfs}, and the parity operator for the entire Hilbert space is given by
\begin{equation}
\hat{\Pi} \equiv \int \ket{-q'}\bra{q'}\ \mathcal{D}[q'] .
\label{parop1}
\end{equation}
It thus follows that one can compute the Wigner functional of an operator by computing its trace with the displaced parity operator, similar to the way it is done without the spatiotemporal degrees of freedom \cite{Royer_1977}.

\subsection{\label{weyl}Functional Weyl transformation}

The inverse process whereby a density functional in either of the quadrature bases is reproduced from its Wigner functional is given by a generalization of the Weyl transformation. For the $q$-basis, we have
\begin{equation}
\rho[q,q'] = \int W\left[\tfrac{1}{2}(q+q'),p\right] \exp[i p\diamond(q-q')]\ \Dcirc [p] .
\end{equation}
A similar expression applies for the $p$-basis. The functional Weyl transformation can also be used to reproduce the density operator:
\begin{equation}
\hat{\rho} = \int \ket{q+\tfrac{1}{2}x} W[q,p] \exp(i p\diamond x) \bra{q-\tfrac{1}{2}x}\ \Dcirc [p]\ \mathcal{D}[q,x] .
\label{weyldigt}
\end{equation}
It then follows that the trace of the density operator is represented by the functional integral of the associated Wigner functional
\begin{equation}
\text{tr}\{\hat{\rho}\} = \int W[\alpha]\ \Dcirc [\alpha] = 1 ,
\label{trwig}
\end{equation}
where we used \eqref{kohreim} and \eqref{alfamaat} to express it in terms of $\alpha$'s, instead of $q$'s and $p$'s.

The expectation value of an observable, which is given by the trace of the product of density operator of the state and the operator for the observable, is represented by the functional integral of the product of their Wigner functionals
\begin{equation}
\text{tr}\{\hat{\rho}\hat{O}\} = \int W_{\hat{\rho}}[\alpha] W_{\hat{O}}[\alpha]\ \Dcirc [\alpha] ,
\label{trobs}
\end{equation}
in the same way it is done with Wigner functions without the spatiotemporal degrees of freedom. The Wigner functional for the observable can be computed in the same way that the Wigner functional is computed for the density operator.

\subsection{\label{char}Characteristic functional}

The characteristic functional is the functional Fourier transform of the Wigner functional
\begin{equation}
\chi[\xi,\zeta] = \int \exp(i p\diamond\zeta-i \xi\diamond q) W[q,p]\ \mathcal{D}[q]\ \Dcirc [p] .
\label{chardef}
\end{equation}
The symplectic form of the functional Fourier kernel is due to the simultaneous Fourier transformation of $q$ and $p$, which are already related by a Fourier relationship \eqref{pqftqp}. The Wigner functional is obtained from the characteristic functional via the inverse functional Fourier transform
\begin{equation}
W[q,p] = \int \exp(i\xi\diamond q-i p\diamond\zeta) \chi[\xi,\zeta]\ \mathcal{D}[\zeta]\ \Dcirc [\xi] .
\label{charnaw}
\end{equation}
The characteristic functional of an operator is directly given by
\begin{equation}
\chi[\xi,\zeta] = \int \bra{q+\tfrac{1}{2}\zeta} \hat{\rho} \ket{q-\tfrac{1}{2}\zeta} \exp(-i \xi\diamond q)\ \mathcal{D}[q] .
\label{char0def}
\end{equation}

The calculation of a characteristic functional can be converted into a purely operator-based calculation, similar to the way it is done for Wigner functionals, by using the shift operators defined in \eqref{skuifqp} and the Baker-Campbell-Hausdorff. The result
\begin{equation}
\chi[\xi,\zeta] = \text{tr} \left\{ \hat{\rho} \hat{D}^{\dag}[\zeta,\xi] \right\} ,
\label{chartr}
\end{equation}
shows that the characteristic functional of an operator is given by its trace with the adjoint of the displacement operator. Note the interchange in the roles of $\xi$ and $\zeta$.

The characteristic functional serves as a generating functional for the moments of the Wigner functional, in the same way it does for Wigner functions without the spatiotemporal degrees of freedom. However, here the derivatives are replaced by functional derivatives. The $(m,n)$-th moment of the Wigner functional is obtain by the $(m,n)$-th functional derivatives with respect to the respective functional variables
\begin{equation}
\int q^m p^n W[q,p]\ \mathcal{D}[q]\ \Dcirc [p] = \left. (i)^m(-i)^n\delta_{\xi}^m\delta_{\zeta}^n\chi[\xi,\zeta]\right|_{\xi=\zeta=0} .
\label{moments}
\end{equation}
Here, $q^m p^n$ is the Wigner functional of the operator with the equivalent powers of quadrature operators in symmetrized order. The functional expression in \eqref{moments} was previously shown in the context of quantum field theory for the spatial degrees of freedom only \cite{Mrowczynski_Wigner_1994}.

\subsection{\label{wigprod}Moyal product or star product}

The Wigner functional for the product of two operators can be obtained by expressing these operators in terms of Weyl transformations \eqref{weyldigt}. The result is a functional integral over the Wigner functionals of these operators:
\begin{align}
W_{\hat{A}\hat{B}}[q,p] = & 2^{2\Omega} \int W_{\hat{A}}\left[q-q_1,p-p_1\right] W_{\hat{B}}\left[q-q_2,p-p_2\right] \nonumber \\
 & \times \exp[ i 2 (q_1\diamond p_2- q_2\diamond p_1)]\ \mathcal{D}[q_1,q_2]\ \Dcirc[p_1,p_2] .
\label{wig2prodop}
\end{align}
It represents the {\em Moyal product} or {\em star product} for Wigner functionals \cite{psqm}. An alternative way to express the star product is in terms of functional derivatives
\begin{equation}
W_{\hat{A}\hat{B}}[q,p] = W_{\hat{A}}[q,p] \exp\left[ \frac{i}{2} \left( \overleftarrow{\delta_q} \overrightarrow{\delta_p} - \overrightarrow{\delta_q} \overleftarrow{\delta_p} \right) \right] W_{\hat{B}}[q,p] ,
\label{wig2proddif}
\end{equation}
where $\delta_q$ and $\delta_p$ represent functional derivatives with respect to $q$ and $p$, respectively, and the arrows indicate to which side the derivatives are applied. One can use the expressions of the Wigner functionals in terms of their characteristic functionals, given in \eqref{charnaw}, to show that \eqref{wig2proddif} leads to \eqref{wig2prodop}.

For the product of three operators, the functional integral expression is
\begin{align}
W_{\hat{A}\hat{B}\hat{C}}[\alpha] = & \int \exp[(\alpha^*-\alpha_a^*)\diamond\alpha_b -\alpha_b^*\diamond(\alpha-\alpha_a)] W_{\hat{A}}\left[\tfrac{1}{2}(\alpha_a+\alpha_b+\alpha)\right] \nonumber \\
& \times W_{\hat{B}}\left[\alpha_a\right] W_{\hat{C}}\left[\tfrac{1}{2}(\alpha_a-\alpha_b+\alpha)\right]\ \Dcirc[\alpha_a,\alpha_b] .
\label{wig3prod}
\end{align}
Here, we used \eqref{kohreim} and \eqref{alfamaat} to render the expression in terms of $\alpha$'s.

\subsection{\label{wigprob}Marginal distributions}

As a quasi-probability distribution over the functional phase space, the Wigner functional of a state does not qualify as a true probability density. However, one can compute a probability density from it by integrating over either $p$ or $q$ (or any linear combination of $p$ and $q$). Integrating the Wigner functional over $p$, we obtain
\begin{align}
\int W[q,p]\ \Dcirc [p] = & \int \bra{q+\tfrac{1}{2}q'}\hat{\rho}\ket{q-\tfrac{1}{2}q'} \delta[q']\ \mathcal{D}[q'] \nonumber \\
= & \bra{q}\hat{\rho}\ket{q} = \rho[q,q] .
\label{propwigq}
\end{align}
Hence, we recover the diagonal of the density functional, which represents the probabilities. To perform the integration over $q$, we first need to insert identities resolved in the $p$-basis
\begin{align}
\int W[q,p]\ \mathcal{D}[q] = & \int \braketa{q+\tfrac{1}{2}q'}{p_1} \bra{p_1}\hat{\rho}
\ket{p_2}\braketb{p_2}{q-\tfrac{1}{2}q'} \exp(-i p\diamond q')\ \Dcirc [p_1,p_2]\ \mathcal{D}[q,q'] \nonumber \\
= & \bra{p}\hat{\rho}\ket{p} = \rho[p,p] .
\label{propwigp}
\end{align}
If we integrate these probability distributions over the remaining variables, we obtain 1, thanks to the normalization.

\subsection{\label{genquasi}General quasi-probability distributions}

Although the spatiotemporal quadrature bases naturally lead to a Wigner functional formulation, other types of quasi-probability distributions can also be used in the formalism. The functional versions of the Glauber-Sudarshan $P$-distribution and the Husimi $Q$-distribution, which are of significant interest in quantum optics, are respectively defined by the following expressions, incorporating fixed-spectrum coherent states:
\begin{equation}
\hat{\rho} = \int \ket{\alpha} P[\alpha] \bra{\alpha}\ \Dcirc[\alpha] ,
\label{pdistal}
\end{equation}
and
\begin{equation}
Q[\alpha] = \bra{\alpha}\hat{\rho}\ket{\alpha} .
\label{qdistal}
\end{equation}

To show how these functional quasi-probability distributions are related to the Wigner functional formulation, we use the Baker-Campbell-Hausdorff formula in \eqref{bch0} to define different versions of the displacement operator, based on their operator ordering:
\begin{equation}
\hat{D}_s[\eta] \equiv \exp\left(\tfrac{1}{2} s \|\eta\|^{2}\right) \hat{D}[\eta]
= \left\{ \begin{array}{lcl}
\exp\left(\hat{a}^{\dag}\diamond\eta\right) \exp\left(-\eta^*\diamond\hat{a}\right) & \text{for} & s=1 \\
\exp\left(\hat{a}^{\dag}\diamond\eta-\eta^*\diamond\hat{a}\right) & \text{for} & s=0 \\
\exp\left(-\eta^*\diamond\hat{a}\right)\exp\left(\hat{a}^{\dag}\diamond\eta\right) & \text{for} & s=-1 \\
\end{array} \right. .
\end{equation}
So, for $s=1$, the operator is normal ordered and for $s=-1$ it is anti-normal ordered. The usual displacement operator in symmetrical order is obtained for $s=0$.

The different versions of the displacement operator can now be used, in analogy with \eqref{chartr}, to compute the associated characteristic functional for a density operator $\hat{\rho}$:
\begin{equation}
\chi_{s}[\xi,\zeta] = \text{tr}\left\{\hat{\rho} \hat{D}_{s}^{\dag}[\eta]\right\} ,
\end{equation}
where $\eta=(\zeta+i\xi)/\sqrt{2}$. The functional Fourier transform of the characteristic functional (assuming it is well-defined), as given in \eqref{charnaw}, then leads to the associated functional quasi-probability distribution $W_{s}[\alpha]$. Hence,
\begin{align}
W_{s}[\alpha] = \text{tr} \left\{ \hat{\rho} \hat{T}_s[\alpha] \right\} ,
\label{genichar}
\end{align}
where
\begin{equation}
\hat{T}_s[\alpha] \equiv \int \hat{D}_s^{\dag}[\eta] \exp(\alpha^*\diamond\eta-\eta^*\diamond\alpha)\ \Dcirc[\eta] .
\label{ftdistal}
\end{equation}
One can now use \eqref{genichar} and \eqref{ftdistal}, together with either \eqref{pdistal} or \eqref{qdistal} to show that $W_{1}[\alpha]=P[\alpha]$ and $W_{-1}[\alpha]=Q[\alpha]$. In the latter case, one can insert an identity resolved in terms of the coherent states \eqref{volkoh} to aid the calculation. The association between the operator ordering and the type of functional quasi-probability distributions is the same as in the case without the spatiotemporal degrees of freedom \cite{PhysRev.177.1882}.

\section{\label{voorb}Examples of Wigner functionals}

\subsection{Fixed-spectrum coherent state}

We substitute $\hat{\rho} \rightarrow \ket{\alpha_F}\bra{\alpha_F}$ into \eqref{wigdef} to obtain the Wigner functional for a fixed-spectrum coherent state,
\begin{equation}
W[q,p] = \int \braketa{q+\tfrac{1}{2}x}{\alpha_F} \braketb{\alpha_F}{q-\tfrac{1}{2}x} \exp(-i p\diamond x)\ \mathcal{D}[x] .
\label{wigpure}
\end{equation}
The expressions for the two overlaps are obtained from \eqref{oorvqkoh0}. After substituting them into \eqref{wigpure} and evaluating the functional integral over $x$, we obtain
\begin{equation}
W[\alpha] = \mathcal{N}_0 \exp\left(-2\|\alpha-\alpha_0\|^2\right) ,
\label{wigpurekoh}
\end{equation}
where $\mathcal{N}_0$ is a normalization constant, and $\alpha_0(\mathbf{k})$ is the parameter function of the fixed-spectrum coherent state. We emphasize that, while the expression looks familiar, one must remember that it represents a functional --- the argument contains an integral of functions over all wave vectors.

The normalization constant $\mathcal{N}_0$ can be obtained either by keeping track of the constants during the calculation or by imposing the requirement that the state is normalized, as in \eqref{trwig}. Both ways lead to
\begin{equation}
\mathcal{N}_0 = 2^{\Omega} .
\label{kohwignorm}
\end{equation}

\subsection{\label{wigcohass}Coherent state assisted calculation}

It is often convenient to employ coherent states in the computation of the Wigner functionals. Inserting identity operators, resolved in terms of coherent states \eqref{volkoh}, into \eqref{wigdef}, we obtain
\begin{equation}
W_{\hat{A}}[q,p] = \int \braketa{q+\tfrac{1}{2}x}{\alpha_1} \bra{\alpha_1}\hat{A}\ket{\alpha_2} \braketb{\alpha_2}{q-\tfrac{1}{2}x} \exp(-i p\diamond x)\ \mathcal{D}[x]\ \Dcirc[\alpha_1,\alpha_2] ,
\end{equation}
where $\hat{A}$ is an arbitrary operator, and $\alpha_1$ and $\alpha_2$ are the parameter functions for with the fixed-spectrum coherent states, serving as integration variables. Next, we substitute \eqref{oorvqkoh0} into the result and evaluate the functional integral over $x$ to obtain
\begin{align}
W_{\hat{A}}[\alpha] = & \mathcal{N}_0 \int \exp\left(-2\|\alpha\|^2+2\alpha^*\diamond\alpha_1+2\alpha_2^*\diamond\alpha
 -\tfrac{1}{2}\|\alpha_1\|^2-\tfrac{1}{2}\|\alpha_2\|^2-\alpha_2^*\diamond\alpha_1 \right) \nonumber \\
& \times \bra{\alpha_1}\hat{A}\ket{\alpha_2}\ \Dcirc[\alpha_1,\alpha_2] ,
\label{asscohwig}
\end{align}
where $\mathcal{N}_0$ is given in \eqref{kohwignorm}. It now remains to evaluate the overlap of the operator $\hat{A}$ by the two coherent states and to perform the functional integrations over $\alpha_1$ and $\alpha_2$ to obtain the Wigner functional for $\hat{A}$.

\subsection{Fixed-spectrum Fock states}

Next, we use the coherent state assisted approach to compute the Wigner functionals for the fixed-spectrum Fock states, defined in \eqref{fsfsdef}. The overlap between such a Fock state and two arbitrary fixed-spectrum coherent states gives
\begin{equation}
\braket{\alpha_1}{n_F} \braket{n_F}{\alpha_2} = \exp \left(-\tfrac{1}{2}\|\alpha_1\|^2-\tfrac{1}{2}\|\alpha_2\|^2\right) \frac{1}{n!} (\langle \alpha_1,F \rangle \langle F,\alpha_2 \rangle)^n ,
\end{equation}
where we used \eqref{fsfsdef}, \eqref{fsskepabs}, \eqref{akopkohm} and \eqref{fsinvac}. One can simplify the calculation by representing the above result as a generating functional
\begin{equation}
\mathcal{K} \equiv \exp\left(-\tfrac{1}{2}\|\alpha_1\|^2-\tfrac{1}{2}\|\alpha_2\|^2
+ \eta_1 \langle\alpha_1,F\rangle + \eta_2 \langle F,\alpha_2 \rangle \right) ,
\end{equation}
where $\eta_1$ and $\eta_1$ are auxiliary parameters, such that
\begin{equation}
\braket{\alpha_1}{n_F}\braket{n_F}{\alpha_2} = \left.\frac{1}{n!}\frac{\partial^n}{\partial\eta_1^n}\frac{\partial^n}{\partial\eta_2^n}\mathcal{K}\right|_{\eta_1=\eta_2=0} .
\label{gen2oper}
\end{equation}
Substituting $\bra{\alpha_1}\hat{A}\ket{\alpha_2}\rightarrow \mathcal{K}$ into \eqref{asscohwig}, we obtain a generating functional for the Wigner functionals of the Fock states, expressed as a functional integral
\begin{align}
\mathcal{W}(\eta) = & \sum_n \eta^n W_{\ket{n}}[\alpha] \nonumber \\
 = & \mathcal{N}_0 \int \exp\left(-2\|\alpha\|^2+2\alpha^*\diamond\alpha_1+2\alpha_2^*\diamond\alpha -\|\alpha_1\|^2 -\|\alpha_2\|^2 \right. \nonumber \\
& \left. -\alpha_2^*\diamond\alpha_1 + \eta_1\alpha_1^*\diamond F+\eta_2 F^*\diamond\alpha_2\right)\ \Dcirc[\alpha_1,\alpha_2] .
\label{fockwig}
\end{align}
The functional integration over $\alpha_1$ and $\alpha_2$ produces
\begin{equation}
\mathcal{W}(\eta_1,\eta_2) = \mathcal{N}_0 \exp\left(-2\|\alpha\|^2 +2\eta_1\alpha^*\diamond F+2\eta_2 F^*\diamond\alpha-\eta_1\eta_2\right) .
\label{fockwig3}
\end{equation}
One can show that the Wigner functionals for the individual fixed-spectrum Fock states are given by
\begin{equation}
W_{\ket{n}}[\alpha] = (-1)^n \mathcal{N}_0 L_n\left(4|\langle F,\alpha\rangle|^2\right)\exp\left( -2|\alpha|^2\right) ,
\label{fockwig4}
\end{equation}
where $L_n$ is the $n$-th order Laguerre polynomial. Although the Wigner functionals for the individual fixed-spectrum Fock states are not in Gaussian form, the generating functional for the Wigner functionals of these Fock states, given in \eqref{fockwig3}, is in Gaussian form and can therefore be used in functional integrals.

\subsection{Wigner functional for the number operator}

Wigner functionals are not only associated with quantum states --- they can also represent arbitrary operators. We now use the coherent state assisted approach to obtain a Wigner functional for the number operator defined in \eqref{numopk}. For this purpose, we compute
\begin{equation}
\bra{\alpha_1}\hat{n}\ket{\alpha_2} = \langle\alpha_1,\alpha_2\rangle
\exp \left(-\tfrac{1}{2}\|\alpha_1\|^2-\tfrac{1}{2}\|\alpha_2\|^2 +\langle\alpha_1,\alpha_2\rangle \right) ,
\end{equation}
where we used \eqref{fsinprodf}. We again use a generating functional to simplify calculations. The generating functional
\begin{equation}
\mathcal{G} = \exp \left(-\tfrac{1}{2}\|\alpha_1\|^2-\tfrac{1}{2}\|\alpha_2|^2+J \langle\alpha_1,\alpha_2\rangle \right) ,
\end{equation}
reproduces the overlap through
\begin{equation}
\bra{\alpha_1}\hat{n}\ket{\alpha_2} = \left. \partial_J \mathcal{G} \right|_{J=1} .
\end{equation}
We substitute $\bra{\alpha_1}\hat{A}\ket{\alpha_2}\rightarrow \mathcal{G}$ into \eqref{asscohwig} and evaluate the functional integrals to obtain
\begin{align}
\mathcal{W}(J) = & \mathcal{N}_0 \int \exp\left(- 2\|\alpha\|^2+2\alpha^*\diamond\alpha_1-\|\alpha_1\|^2-\|\alpha_2\|^2 \right. \nonumber \\
& \left. +2\alpha_2^*\diamond\alpha-\alpha_2^*\diamond\alpha_1 +J \alpha_1^*\diamond\alpha_2 \right)\ \Dcirc[\alpha_1,\alpha_2] \nonumber \\
 = & \frac{\mathcal{N}_0}{(1+J)^{\Omega}} \exp\left[- 2\left(\frac{1-J}{1+J}\right)\alpha^*\diamond\alpha\right] .
\label{numwig}
\end{align}
Finally, we evaluate the derivative with respect to $J$ and set $J=1$ to get the Wigner functional for the number operator:
\begin{equation}
W_{\hat{n}}[\alpha] = \alpha^*\diamond\alpha - \tfrac{1}{2}\Omega .
\label{wignum1}
\end{equation}

\subsection{\label{wigdisp}Displacement operator}

Next, we consider the displacement operator given in \eqref{verplfs}. Using the Baker-Campbell-Hausdorff formula
\begin{equation}
\exp\left(\hat{X}+\hat{Y}\right) = \exp\left(-\tfrac{1}{2} [\hat{X},\hat{Y}] \right) \exp(\hat{X}) \exp(\hat{Y}) ,
\label{bch0}
\end{equation}
which assumes $[\hat{X},[\hat{X},\hat{Y}]]=[\hat{Y},[\hat{X},\hat{Y}]]=0$, to separate the displacement operator into a product of exponential operators, we obtain
\begin{equation}
\hat{D}[\alpha_0]= \exp\left(-\tfrac{1}{2}\|\alpha_0\|^2\right) \exp\left(\alpha\diamond\hat{a}^{\dag}\right)
\exp\left(-\alpha^*\diamond\hat{a}\right) .
\end{equation}
It allows us to compute the overlap
\begin{align}
\bra{\alpha_1} \hat{D}[\alpha_0]\ket{\alpha_2} = & \exp \left(-\tfrac{1}{2} \|\alpha_0\|^2+\alpha_1^*\diamond\alpha_0
-\tfrac{1}{2}\|\alpha_1\|^2 \right. \nonumber \\
& \left. -\tfrac{1}{2}\|\alpha_2\|^2+\alpha_1^*\diamond\alpha_2-\alpha_0^*\diamond\alpha_2 \right) .
\end{align}
for the coherent state assisted approach. After substituting it into \eqref{asscohwig}, we obtain
\begin{align}
W_{\hat{D}} = & \mathcal{N}_0 \int \exp\left(- 2\|\alpha\|^2+2\alpha^*\diamond\alpha_1+2\alpha_2^*\diamond\alpha
 -\|\alpha_1\|^2-\|\alpha_2\|^2 \right. \nonumber \\
 & \left. -\alpha_2^*\diamond\alpha_1+\alpha_1^*\diamond\alpha_0+\alpha_1^*\diamond\alpha_2
 -\alpha_0^*\diamond\alpha_2-\tfrac{1}{2} \|\alpha_0\|^2 \right)\ \Dcirc[\alpha_1,\alpha_2] .
\label{verpwig0}
\end{align}
Finally, we evaluate the functional integrals over $\alpha_1$ and $\alpha_2$ and obtain a familiar form:
\begin{equation}
W_{\hat{D}}[\alpha;\alpha_0] = \exp\left(\alpha^*\diamond\alpha_0-\alpha_0^*\diamond\alpha \right) .
\label{verpwig1}
\end{equation}

One can use the Wigner functional expression for the product of three operators in \eqref{wig3prod} to obtain a general expression for the Wigner functional of an arbitrary state after displacement operators are applied to it:
\begin{align}
W_{\hat{D}\hat{\rho}\hat{D}^{\dag}}[\alpha] = & \int \exp\left(\alpha_b^*\diamond\alpha_0-\alpha_0^*\diamond\alpha_b\right)
 \exp[(\alpha^*-\alpha_a^*)\diamond\alpha_b-\alpha_b^*\diamond(\alpha-\alpha_a)] \nonumber \\
 & \times W_{\hat{\rho}}\left[\alpha_a\right]\ \Dcirc[\alpha_a,\alpha_b] .
\end{align}
There are no quadratic terms for $\alpha_b$ in the exponent. Hence, the functional integration over $\alpha_b$ produces a Dirac $\delta$ functional, as in \eqref{adirac}, which leads to
\begin{equation}
W_{\hat{D}\hat{\rho}\hat{D}^{\dag}}[\alpha] = (2\pi)^{\Omega} \int W_{\hat{\rho}}\left[\alpha_a\right]
\delta[\alpha_a-\alpha+\alpha_0]\ \Dcirc[\alpha_a]
 = W_{\hat{\rho}}\left[\alpha-\alpha_0\right] .
\end{equation}
As expected, the effect of the displacement operation on an arbitrary Wigner functional is a shift in its argument.

\section{\label{disc}Discussion}

The notation for the formalism that we present here leads to expressions that are almost identical to those that exclude the spatiotemporal degrees of freedom. While the notation may alleviate the complexities in its use, this resemblance could be misleading --- causing one to confuse the more general expressions for those of the simpler case. Therefore, it is necessary to emphasize the difference in meaning.

Consider for example the Wigner function of the coherent state incorporating only the particle-number degrees of freedom, given by
\begin{equation}
W(\alpha;\alpha_0) = 2 \exp\left(-2|\alpha-\alpha_0|^2 \right) ,
\label{verg1}
\end{equation}
and compare it to the Wigner functional of the fixed-spectrum coherent state, which incorporates both the spatiotemporal and particle-number degrees of freedom
\begin{equation}
W[\alpha;\alpha_0] = 2^{\Omega} \exp\left(-2\|\alpha-\alpha_0\|^2 \right) .
\label{verg2}
\end{equation}
While the two expressions may seem very similar, they represent different content. In \eqref{verg1}, $\alpha$ and $\alpha_0$ respectively represent a complex variable and a complex parameter. On the other hand, in \eqref{verg2} they represent {\em functions}. That is why the former is a Wigner function, while the latter is a Wigner functional. The argument of the former contains the modulus square of the difference between the complex values. In contrast, the argument of the latter contains an integral over the space of wave vectors that computes the squared magnitude of the difference between the complex functions. Due to the integral, the wave vector dependencies of the complex functions are removed (integrated out). As a result, the Wigner functional does not explicitly depend on the wave vector. However, it does depend on the complex function as a whole.

Another difference between the expressions in \eqref{verg1} and \eqref{verg2} is the normalization constant. In the former case, the constant is a finite number. In the latter case, it becomes a divergent constant that one can associate with the cardinality of the space. If $\Omega$ is associated with the cardinality of countable infinity, then $2^{\Omega}$ would represent the cardinality of the continuum. It is inevitable that functional integrals would produce such divergent constants. However, when these functional integrals are employed to compute the predicted results of measurements, one expects to obtain finite quantitative results. Therefore, the divergent constants must cancel. According to cardinal arithmetic \cite{card}, all divergent constants of the same cardinality are formally equal. However, unless one can keep careful track of these constants, their cancellation may hide finite (ordinal) numbers that are important for the correct quantitative predictions. For this reason, we retain the precise form of the divergent constants, be it $\pi^{\Omega/4}$, $2^{\Omega}$, $(2\pi)^{\Omega}$, or whatever else, even though all these constants are formally equal.

One can also compare the current formalism with the continuous variable formalism \cite{weedbrook}. The latter represents the spatiotemporal degrees of freedom as a finite number of discrete modes. The implication is that the complete Hilbert space of quantum optical states is stratified into a tensor product of discrete Hilbert spaces, each representing one mode. Such a stratification cannot completely separate these Hilbert spaces because they share the same vacuum state. Practical calculations require a truncation to get a finite number of such Hilbert spaces. A coherent state in this formalism is expressed as
\begin{equation}
W[\mathbf{Q};\mathbf{Q}_0] = 2^M \exp\left[-2\left(\mathbf{Q}-\mathbf{Q}_0\right)^T J \left(\mathbf{Q}-\mathbf{Q}_0\right)\right] ,
\label{verg3}
\end{equation}
where $\mathbf{Q}$ is a vector consisting of $M$ pairs of quadrature variables, one pair for each of the $M$ Hilbert spaces; $\mathbf{Q}_0$ is a vector of the associated parameters and $J$ is a symplectic matrix that maintains the correct multiplications among the quadrature variables. Although there are many applications where the latter formalism has been used successfully, such as representing the spectral components in a frequency comb \cite{roslund_wavelength-multiplexed_2014}, these cases invariably truncate the set of discrete modes to a finite number. Therefore, such cases do not represent the complete Hilbert space of all quantum optical states.

\section{\label{concl}Conclusion}

The existence of a complete orthogonal basis for the full Hilbert space of quantum optical states, which incorporates all the degrees of freedom associated with photonic states, allows one to formulate powerful tools to analyze quantum optical systems. Since the complete orthogonal basis is a quadrature basis, the natural choice of such a formalism is a generalization of the well-known Wigner distribution.

Here, the Wigner functional formalism is presented, based on the spatiotemporal quadrature bases. The result demonstrates a clear analogy between the functional formalism and the well-known Wigner function, characteristic function and Weyl transform. Even the star product is reproduced in a similar form.

We use the functional formalism to compute some examples of Wigner functionals. These examples include: the Wigner functional for fixed-spectrum coherent states, a generating functional for the Wigner functionals of fixed-spectrum Fock states and the Wigner functionals for the number operator and the displacement operator.

Functional Wigner distributions were previously introduced in the context of quantum field theory to represent bosonic \cite{Mrowczynski_Wigner_1994} and fermionic \cite{Mrowczynski_Wigner_2013} quantum fields. (See also \cite{recent}.) Here, our interest is quantum optics. While we only focus on the (bosonic) optical field, one could also investigate the extension of the current formalism to fermionic fields, which could be applied in solid state physics for instance.

\section*{Acknowledgement}

We gratefully acknowledge fruitful discussions with Thomas Konrad. This work was supported in part by funding from the National Research Foundation of South Africa (Grant Numbers: 118532).




\end{document}